\documentclass{ws-procs975x65}            % default, citations in superscript
\def\figsubcap#1{\par\noindent\centering\footnotesize(#1)}
\usepackage{hyperref}
\begin{document}
\title{Direct photon and neutral pion production in pp and Pb--Pb collisions measured with the ALICE experiment at LHC}

\author{D. Peressounko$^*$ for the ALICE Collaboration}

\address{RRC "Kurchatov institute", \\
Kurchatov sq.,1, Moscow, 123182, Russia\\
$^*$E-mail: Dmitri.Peressounko@cern.ch}

\begin{abstract} 
  Measurements of direct photon and neutral pion production in heavy-ion
  collisions provide a comprehensive set of observables characterizing
  properties of the hot QCD medium. Direct photons provide means to test the
  initial stage of an AA collision and carry information about the temperature
  and space-time evolution of the hot medium. Neutral pion suppression probes
  the parton energy loss in the hot medium. Measurements of neutral meson
  spectra in pp collisions at LHC energies $\sqrt{s}=$0.9, 2.76, 7 TeV serve as
  a reference for heavy-ion collisions and also provide valuable input data for
  parameterization of the QCD parton Fragmentation Functions. In this talk,
  results from the ALICE experiment on direct photon and neutral pion
  production in pp and Pb--Pb collisions are summarized.  
\end{abstract}


\bodymatter

\section{Introduction}

The ultimate goal of heavy ion collisions is a detailed study of the properties
of quark-gluon matter.  Photons provide tools for testing almost all key
features of this hot matter: spectra and correlations of hard hadrons, which
can be reconstructed using photon decays, test the energy loss of energetic partons
in hot matter and the collective flow of identified hadrons up to high $p_{\mathrm
T}$.  {\it Direct} photons, defined as photons not from hadronic decays, are
made up of {\it prompt} photons (emitted by partons of colliding nucleons) and
{\it thermal} photons (emitted by the hot matter analogously to blackbody
radiation).  The former allows us to probe the inital state of the collision
while the latter reflects the temperature and the space-time evolution of the
matter.  In addition, the thermal-photon flow reflects the development of the
hot matter's collective flow at all stages of the collision.

\section{Experimental Setup}
A detailed description of the ALICE experimental setup can be found in
\cite{Aamodt:2008zz}.  Compared to other LHC experiments, ALICE excels at
low-$p_{\mathrm T}$ physics and advanced particle identification. The core of
ALICE is the central tracking system consisting of the Inner Tracking System
and the Time Projection Chamber.  Charged particle identification is improved
by the Transition Radiation Detector and Time Of Flight detector. Finally,
ALICE has two calorimeters, the PHoton Spectrometer (PHOS) and the
electromagetic calorimeter (EMCal), as well as a set of smaller detectors for
triggering and characterizing events. 

\begin{figure}[ht]
  \begin{center}
    \includegraphics[width=0.9\textwidth]{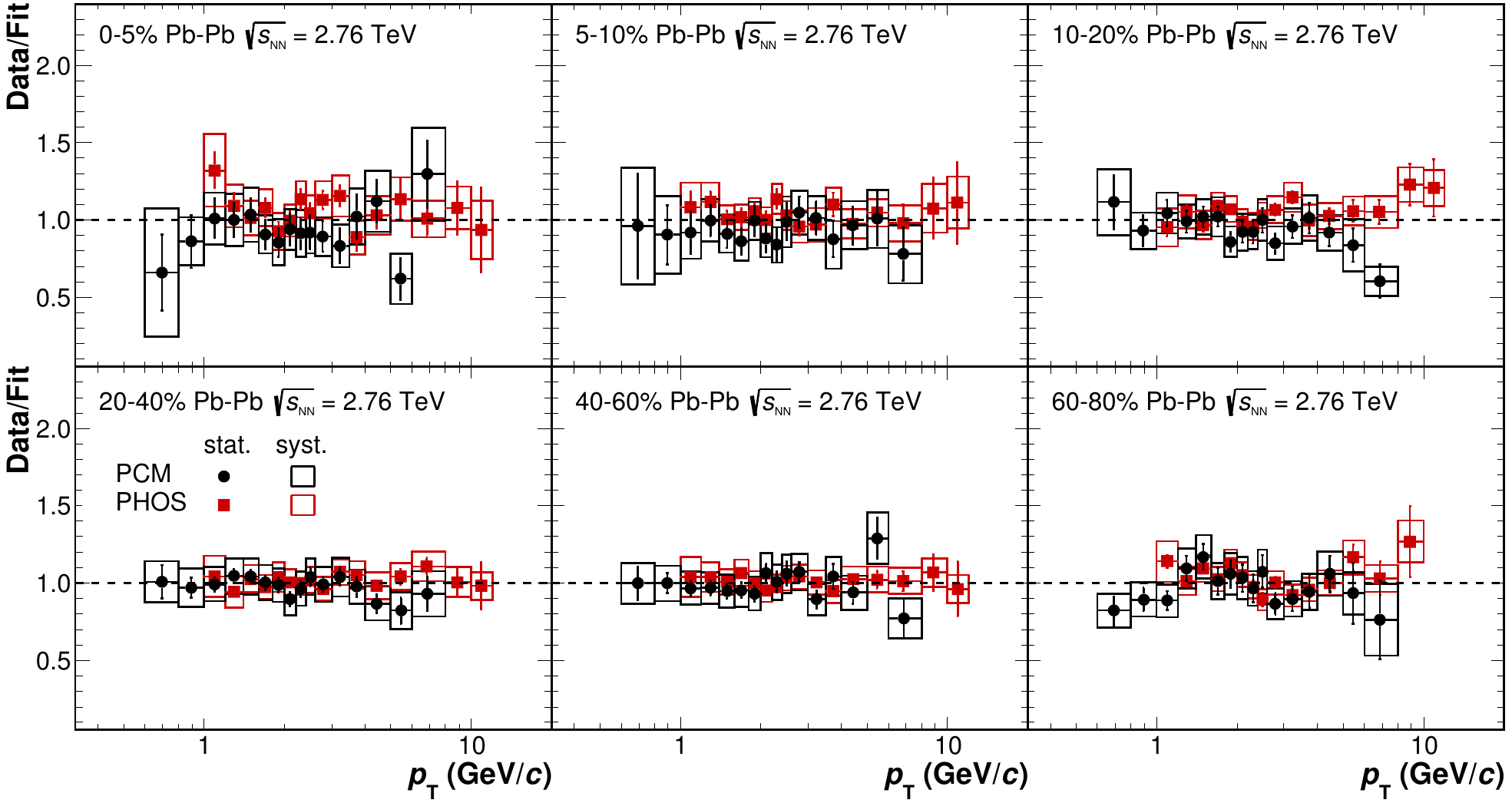}
  \end{center}
  \caption{Comparison of $\pi^0$ spectra measured  in Pb--Pb
    collisions at $\sqrt{s_{NN}}=2.76$ TeV in 6 centrality bins with PHOS and PCM techniques. Both
  spectra are divided by the fit to the combined spectrum \cite{Abelev:2014ypa}.}
  \label{fig:PHOS-PCM-comp}
\end{figure}

\section{Neutral Pion Measurements}
Photons can be reconstructed in ALICE in several ways: using traditional
calorimetry with the PHOS and EMCal or by the Photon Conversion Method (PCM)
via reconstructing  $e^+e^-$ tracks from photons conversion in the central tracking
system \cite{Abelev:2012cn}. PHOS has fine granularity leading to excellent energy and position
resolution though it has a relatively small acceptance. PCM provides good
position and energy resolution and full $2\pi$ coverage in the azimuth.
However, since ALICE was constructed to minimize the material budget, the
photon conversion probability before the middle of the TPC, where tracks still can
be reconstructed with high efficiency, is about 8\%. As a result, both methods
have comparable acceptance~$\times$~efficiencies.

The ability to simultaneously measure photon and neutral pion spectra with
several independent detectors improves the reliability of the final results.
Moreover, the PCM and PHOS measurements have distinct systematic uncertainties,
opposite dependencies of $p_{\mathrm T}$ resolution and different sensitivities
to pileup.  The agreement between the two measurements is shown in fig.\
\ref{fig:PHOS-PCM-comp}, where the $\pi^0$ spectra in different centrality bins
in Pb--Pb collisions at $\sqrt{s_{NN}}=2.76$ TeV are compared.  To elucidate the
comparison both spectra are divided by the fit to the combined spectrum.
Within the statistical and systematic uncertainties the measurements are in
good agreement \cite{Abelev:2014ypa}.

\begin{figure}[h]
  \begin{center}
    \includegraphics[width=0.85\textwidth]{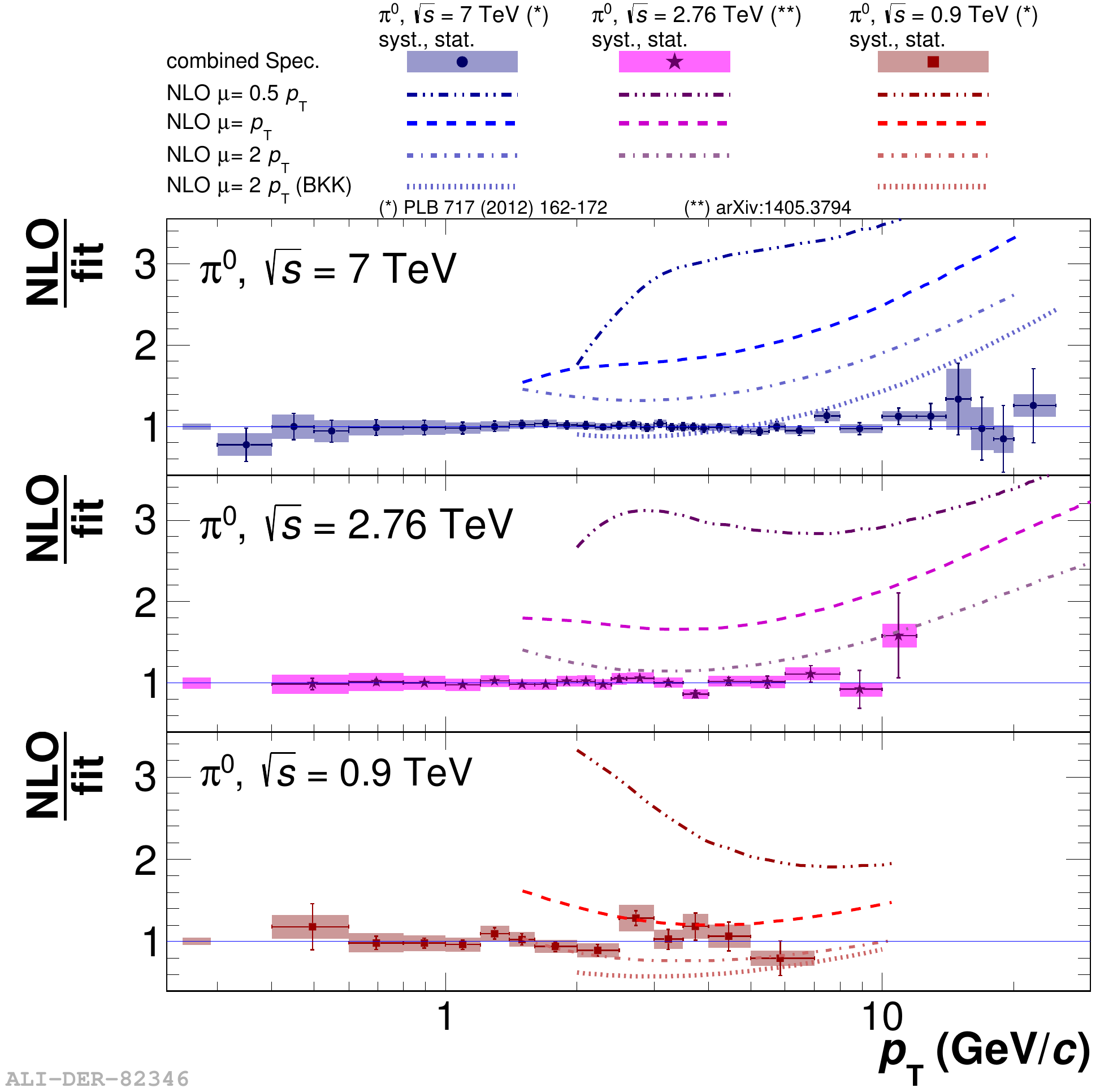}
  \end{center}
  \caption{Comparison of $\pi^0$ spectra measured in pp collisions at
    $\sqrt{s}=0.9,2.76$ and 7~TeV with several pQCD predictions. For clarity both
  data and predictions are divided by the fit to data.}
  \label{fig:Comp-pp-theory}
\end{figure}

We compare the $\pi^0$ spectrum  in pp collisions at $\sqrt{s}=0.9,2.76$ and 7
TeV with several selected pQCD predictions in fig.\ \ref{fig:Comp-pp-theory}.
To make the comparison of steeply falling spectra clear, we fit our spectra to
a function and plot ratios of the measured spectra to the fit along with the
pQCD predictions to the fit \cite{Abelev:2012cn}.  pQCD approximately reproduces the hadron yield at
the lowest LHC energy ($\sqrt{s}=0.9$ TeV) but over-predicts the yield by a
factor of two at higher energies ($\sqrt{s}=2.76, 7$ TeV).  This is typical for
LHC measurements \cite{d'Enterria:2014zga}.  In contrast, pQCD well reproduces
jet spectra in pp collisions at LHC energies \cite{Aad:2011fc}. A possible
explanation is that gluon fragmentation becomes increasingly important, while
the gluon fragmentation functions are not well restricted by existing data at
lower $\sqrt{s}$. This uncertainty should be reduced in the recent calculation
of fragmentation functions \cite{deFlorian:2014xna}, which includes these ALICE
results in their global analysis.

\begin{figure}[h] \begin{center}
  \includegraphics[width=0.9\textwidth]{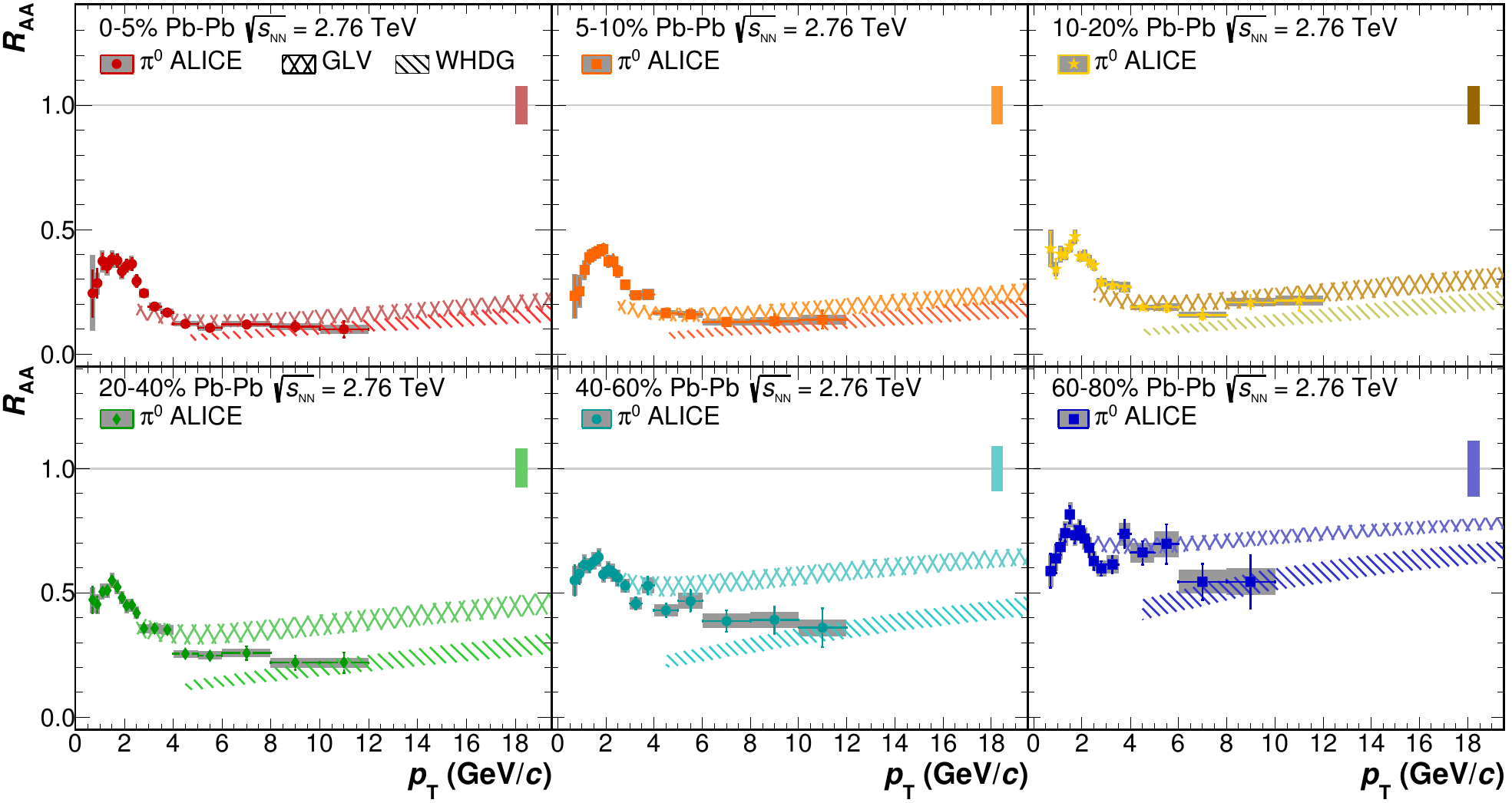}
\end{center} \caption{Neutral pion nuclear modification factor in Pb--Pb
  collisions at $\sqrt{s_{NN}}=2.76$ GeV in 6 centrality classes \cite{Abelev:2014ypa}. For
  comparison, predictions of models of Vitev (GLV)
  \cite{Sharma:2009hn,Neufeld:2010dz} and Horowitz (WHDG) \cite{Horowitz:2011gd} are
  shown.} \label{fig:Raa} 
\end{figure}

ALICE measured neutral pion production in Pb--Pb collisions at 
$\sqrt{s_{NN}}=2.76$~TeV with two techniques: PHOS and PCM. Combined spectra
were produced for 6 centrality classes. For a quantitative estimate of the parton energy
loss, the nuclear modification factor 

\begin{equation}
  R_{AA}(p_{\mathrm T})=\frac{\mathrm{d}^2N/\mathrm{d}p_{\mathrm T} \mathrm{d}y|_\mathrm{AA}}
  {\langle T_{AA} \rangle \times \mathrm{d}^2\sigma/\mathrm{d}p_{\mathrm T} \mathrm{d}y|_\mathrm{pp}}
  \label{eq:raa}
\end{equation}

was calculated, where the nuclear overlap function $\langle T_{AA} \rangle$ is
related to the average number of inelastic nucleon-nucleon collisions $\langle
N_\mathrm{coll} \rangle$  and the pp inelastic cross section
$\sigma_\mathrm{inel}^\mathrm{pp}$ as $\langle T_{AA} \rangle = \langle
N_\mathrm{coll} \rangle / \sigma_\mathrm{inel}^\mathrm{pp}$.  The neutral pion
nuclear modification factors for 6 centrality bins are shown in fig.\
\ref{fig:Raa}.  At $p_{\mathrm T}\gtrsim 5$ GeV/$c$ pions from hard
interactions dominate and the strong suppression in $R_{AA}$ reflects
considerable energy loss for hard partons. At low $p_{\mathrm T}$ ($\lesssim
2$~GeV/$c$) the spectrum is defined by collective (hydrodynamic) expansion and
comparing with pp is not particularly meaningful. The intermediate $p_{\mathrm
T}$ region transitions between these two regimes. In the most central
collisions $R_{AA}$ reaches a minimum of $\sim 0.1$ indicating twice stronger
suppression compared to central Au-Au collisions at RHIC energies
\cite{Adler:2006bw}. Since the spectra of initial partons at LHC energies is
considerably harder, stronger suppression means much larger energy loss
compared to RHIC. 

Fig.\ \ref{fig:Raa} compares the measured nuclear modification factors with
predictions from two advanced models.  Currently, there is no consensus as to
which features are most important and should be incorporated in descriptions of
energy loss in AA collisions. In the Vitev et al.\ model
\cite{Sharma:2009hn,Neufeld:2010dz}, in addition to collisional energy loss,
initial state effects are included, while Horowitz et al.\
\cite{Horowitz:2011gd} account for geometrical fluctuations of hard processes.
Basically, both models are close to data, though Horowitz's is not as good in
reproducing the centrality dependence.

\section{Direct photon spectra and flow}

ALICE measures the direct photon yield via a statistical approach; the
estimated decay photon spectrum is subtracted from the measured inclusive
photon spectrum.  In this analysis one first constructs a double ratio

\begin{equation}
  R_{\gamma}=\frac{\gamma^{inclusive}/\pi^0_{measured}}{\gamma^{decay}/\pi^0_{param}}
  \approx\frac{\gamma^{inclusive}}{\gamma^{decay}}. 
  \label{Rg}
\end{equation}

\begin{figure}[h]%
\begin{center}
  \parbox{0.49\textwidth}{\includegraphics[width=0.49\textwidth]
  {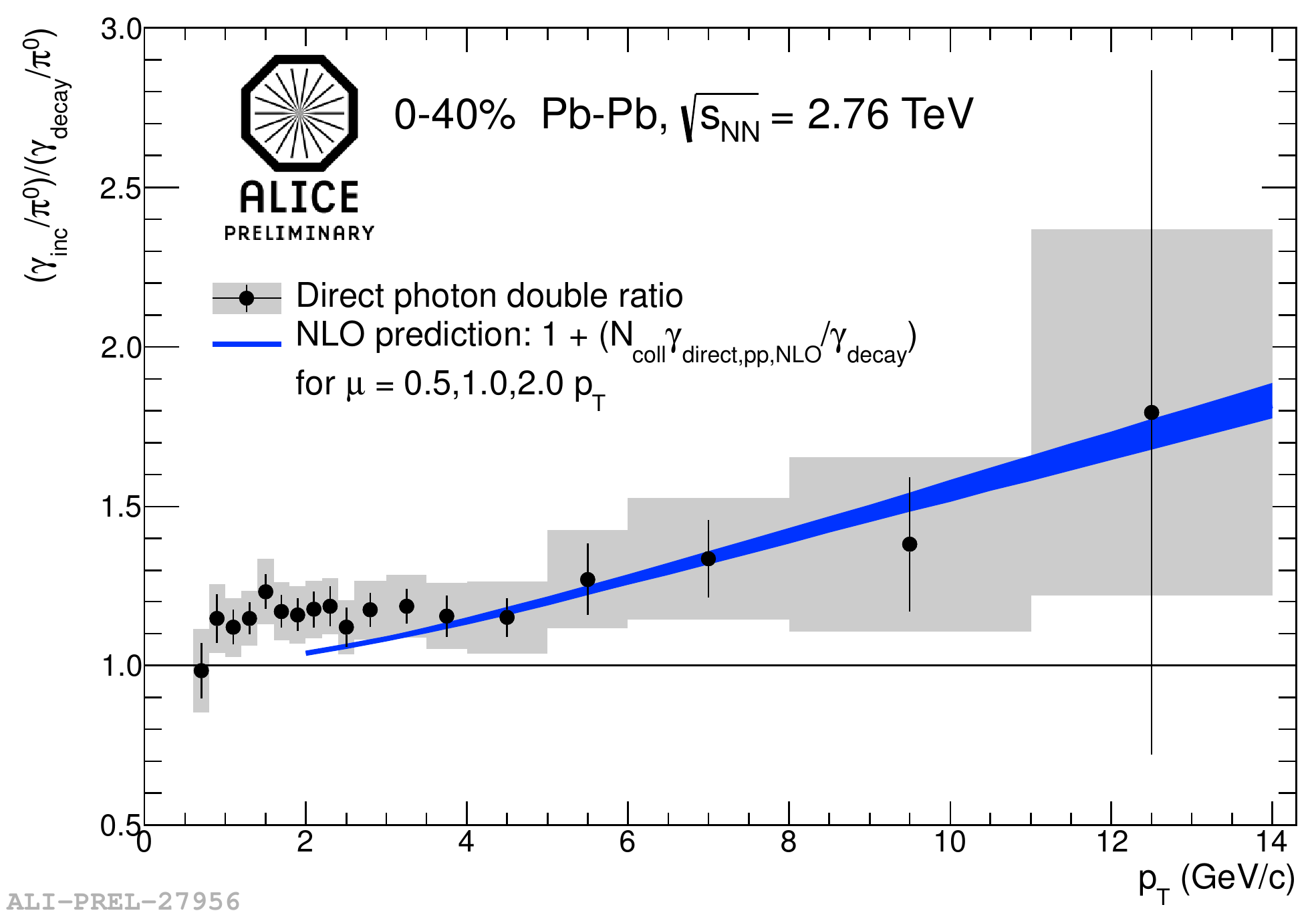}\figsubcap{a}}
  \hfill
  \parbox{0.49\textwidth}{\includegraphics[width=0.49\textwidth]
  {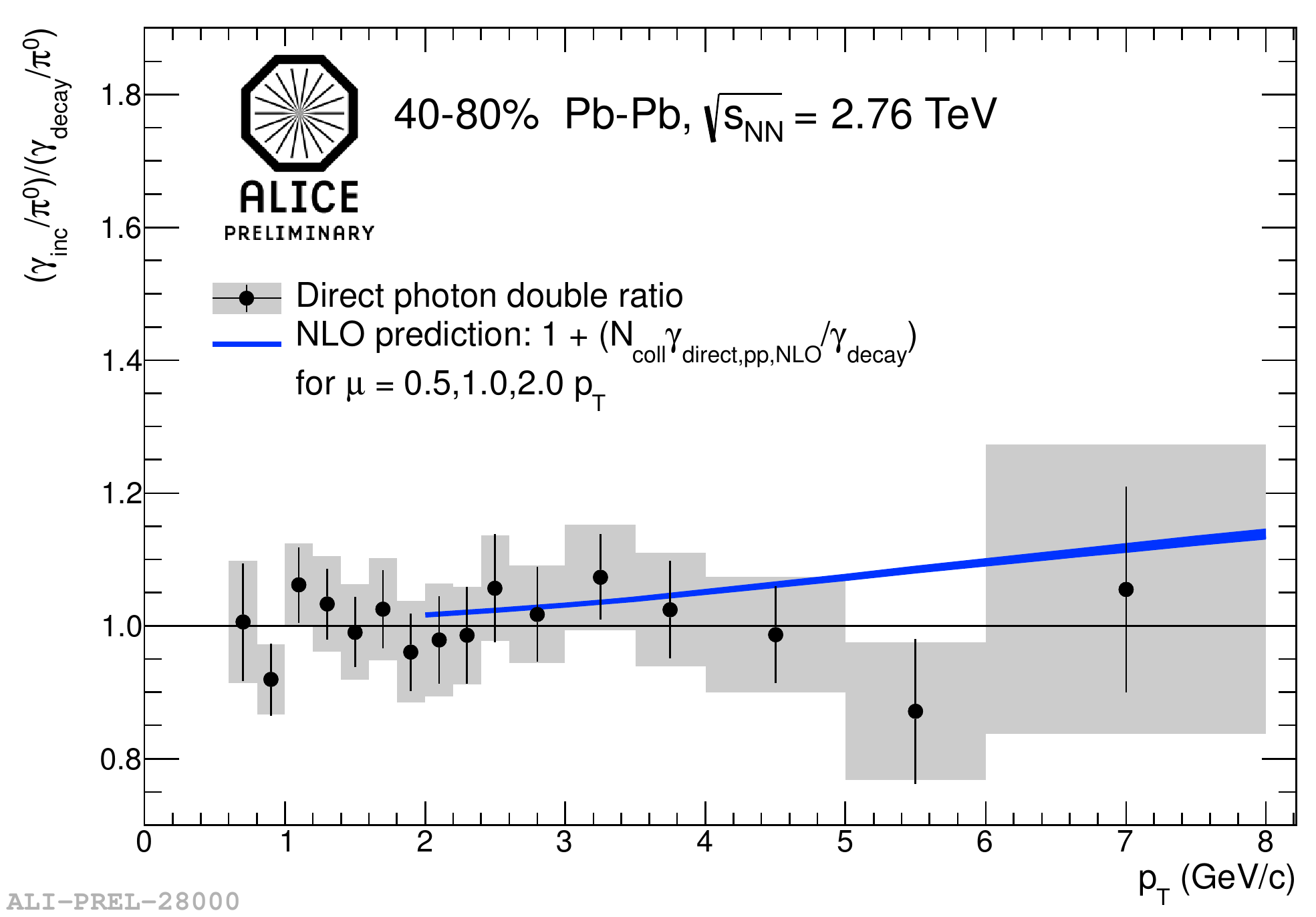}\figsubcap{b}}
  \caption{Photon double ratios in central (a) and peripheral (b) Pb--Pb
  collisions at $\sqrt{s_{NN}}=2.76$~TeV.} 
  \label{fig:Rg}
\end{center}
\end{figure}

A double ratio of unity, $R_{\gamma}=1$, represents no direct photon yield
whereas an increase above unity constitutes a presence of direct photons.  The
advantage of this approach is that most of the largest systematic uncertainties
cancel in the ratio.  Preliminary ALICE results are shown in fig.\
\ref{fig:Rg}(a) for central and fig.\ \ref{fig:Rg}(b) for peripheral events.
Blue bands show a pQCD prediction of the contribution from prompt direct
photons \cite{Gordon:1994ut,Vog97a,Vog04a}. ALICE finds no excess of direct
photons in peripheral events as data agree with pQCD predictions within the
uncertainties. In central collisions, the direct photon yield agrees with the
prompt photon yield at high $p_{\mathrm T}$ but indicate additional excess at
low $p_{\mathrm T}$.

\begin{figure}[h]
  \begin{center}
    \includegraphics[width=0.7\textwidth]{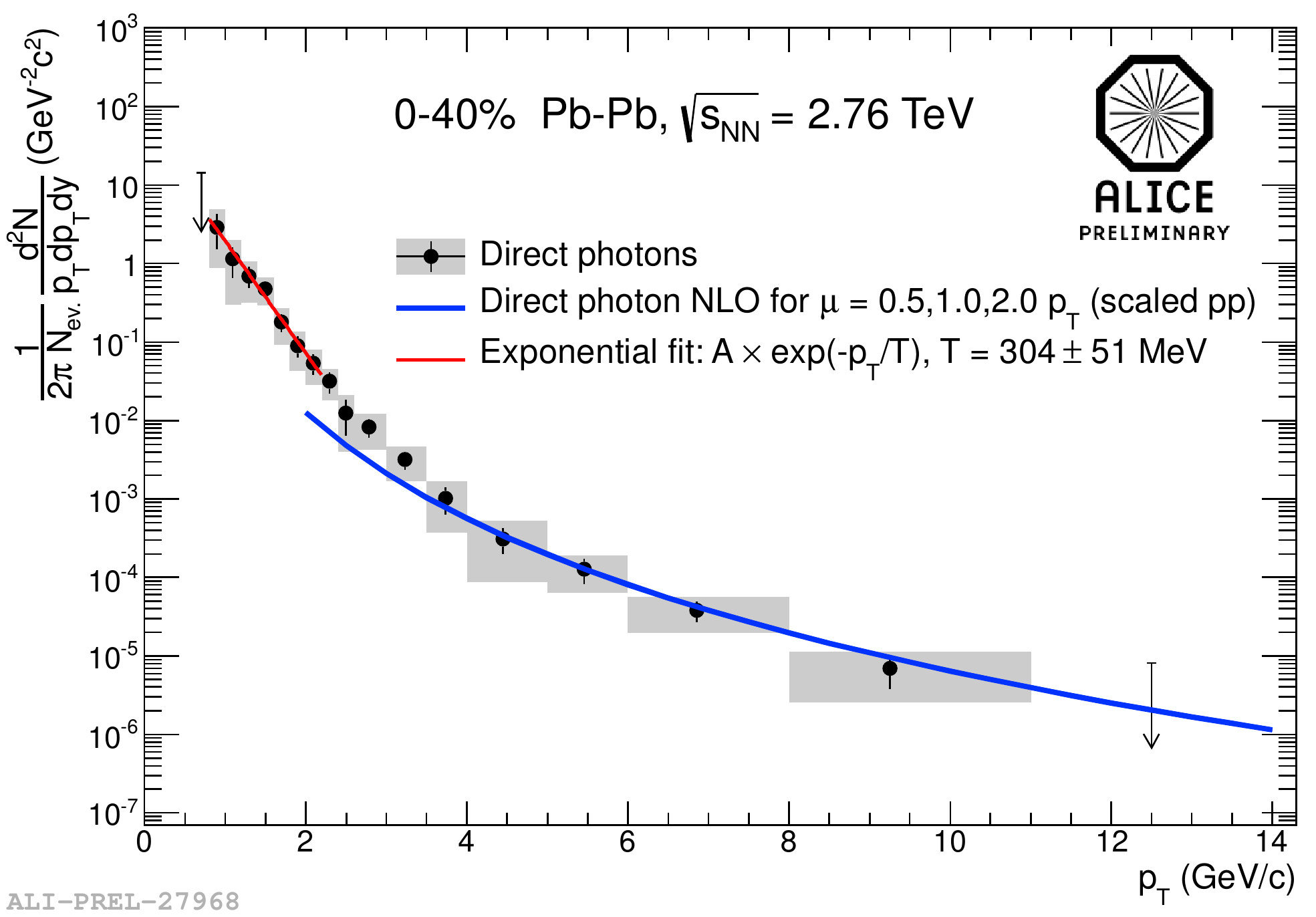}
  \end{center}
  \caption{Direct photon spectrum in central (0-40\%) Pb--Pb collisions at
    $\sqrt{s_{NN}}=2.76$ TeV. The blue line represents pQCD prompt photon
    predictions while the red line is an exponential fit in the range
  $0.9<p_{\mathrm T}<2.0$ GeV/$c$.} 
  \label{fig:gDir}
\end{figure}

From the measured double ratio (fig. \ref{fig:Rg}), the direct photon spectrum is
derived as $N_{\gamma}^{dir}=(1-1/R_{\gamma})N_{\gamma}^{inclusive}$ and
plotted in fig.\ \ref{fig:gDir}. For comparison, pQCD predictions
\cite{Vog97a,Vog04a} are shown by the blue band. In addition, in the region
{$p_{\mathrm T}<2$ GeV/$c$} where one expects a dominant contribution of thermal
direct photons, ALICE fit the data and extracted the inverse slope. However,
one should keep in mind that this slope is influenced by collective expansion
and the entire evolution of the system and therefore should not be directly
interpreted as the temperature in the center of the fireball.

\begin{figure}[h]%
  \begin{center}
%    \parbox{0.52\textwidth}{\includegraphics[width=0.51\textwidth]{figures/2014-May-15-InclPhotonv2_withTheory.pdf}\figsubcap{a}}
    \parbox{0.44\textwidth}{\includegraphics[width=0.44\textwidth]{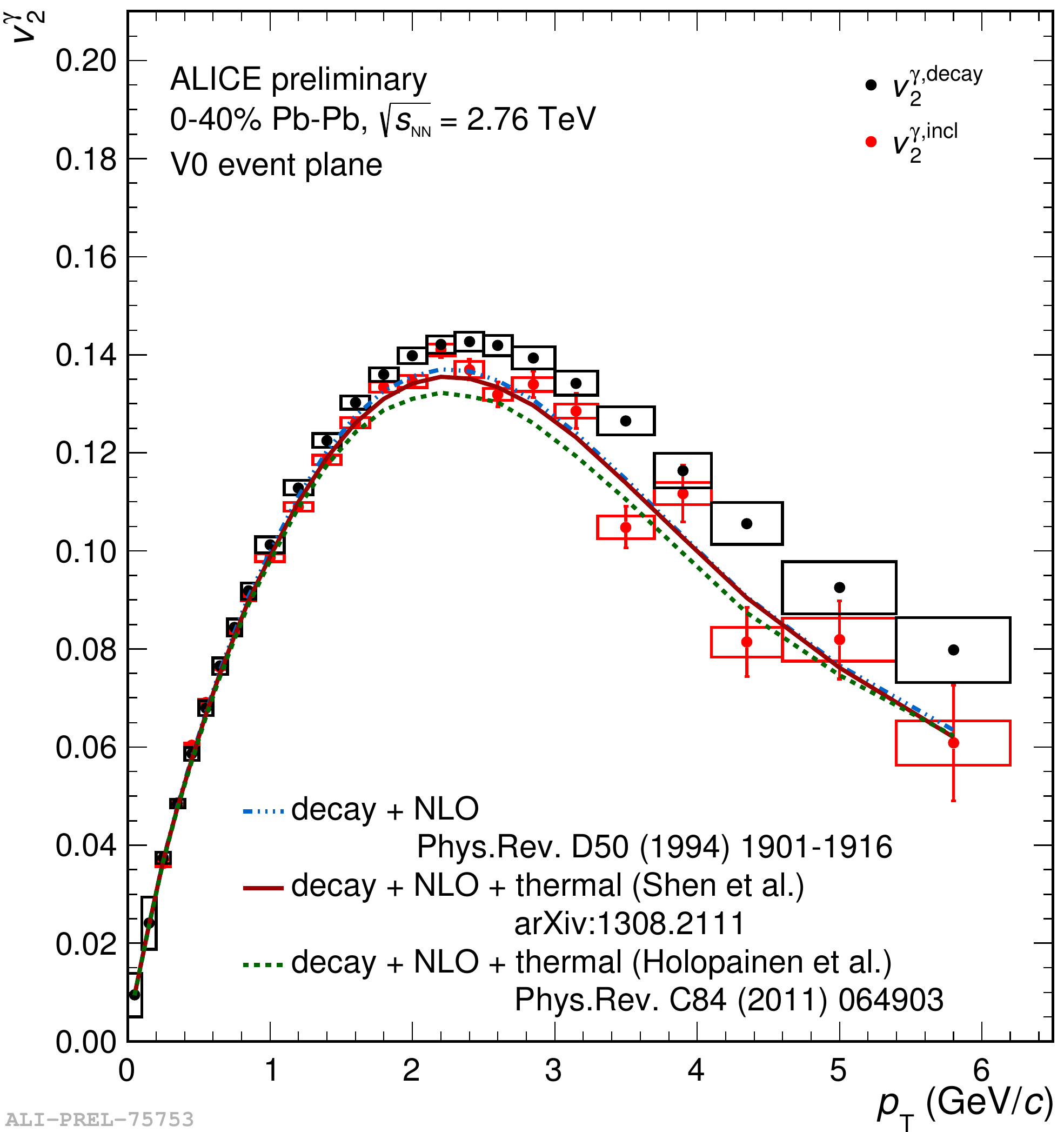}\figsubcap{a}}
    \hfill
%    \parbox{0.41\textwidth}{\includegraphics[width=0.405\textwidth]{figures/2014-May-15-RatiosV2.pdf}\figsubcap{b}}
    \parbox{0.34\textwidth}{\includegraphics[width=0.34\textwidth]{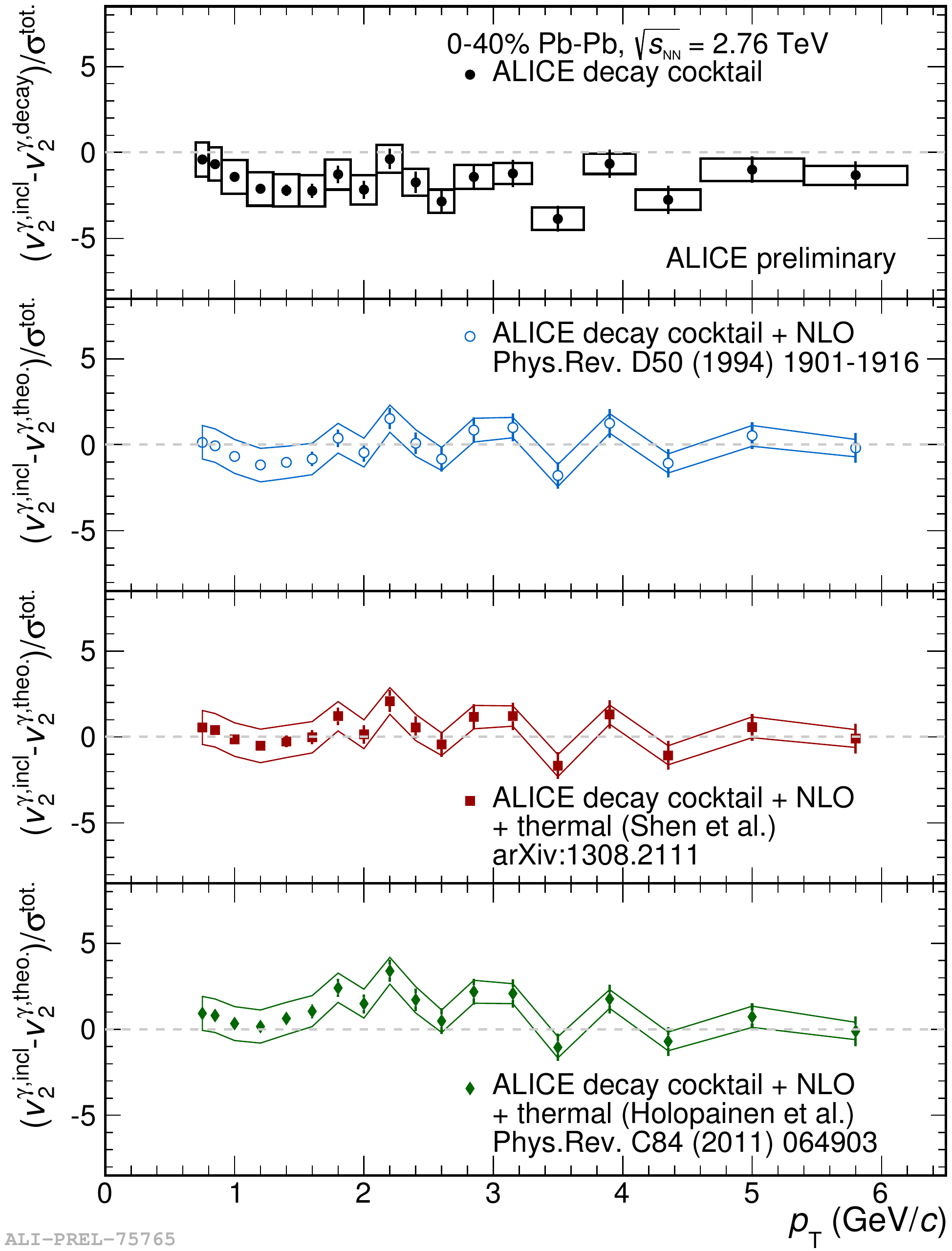}\figsubcap{b}}
    \caption{(a) Comparison of elliptic flow of inclusive and decay photons.
      Lines represent contributions of decay photons with theoretical calculations
      \cite{Gordon:1994ut,Shen:2013cca,Holopainen:2011pd}. (b) Difference between
      inclusive and decay (top plot) and inclusive and theory (3 bottom plots)
    elliptic flows in units of total error.}% 
  \label{fig:v2} \end{center}
\end{figure}

\begin{figure}[h]%
\begin{center}
%  \parbox{0.52\textwidth}{\includegraphics[width=0.51\textwidth]{figures/2014-May-15-InclPhotonv3_withTheory.pdf}\figsubcap{a}}
  \parbox{0.44\textwidth}{\includegraphics[width=0.44\textwidth]{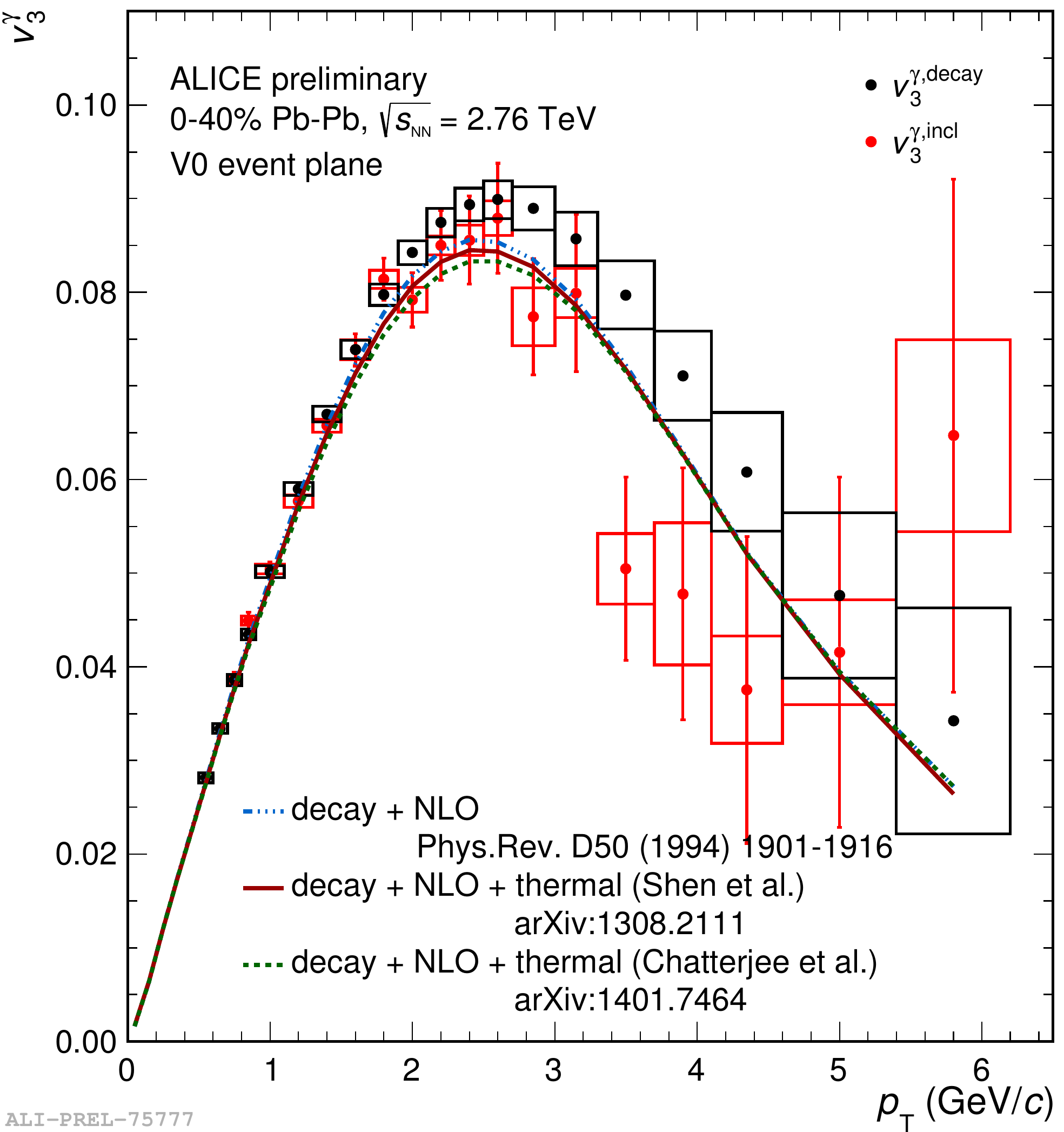}\figsubcap{a}}
  \hfill
%  \parbox{0.41\textwidth}{\includegraphics[width=0.405\textwidth]{figures/2014-May-15-RatiosV3.pdf}\figsubcap{b}}
  \parbox{0.34\textwidth}{\includegraphics[width=0.34\textwidth]{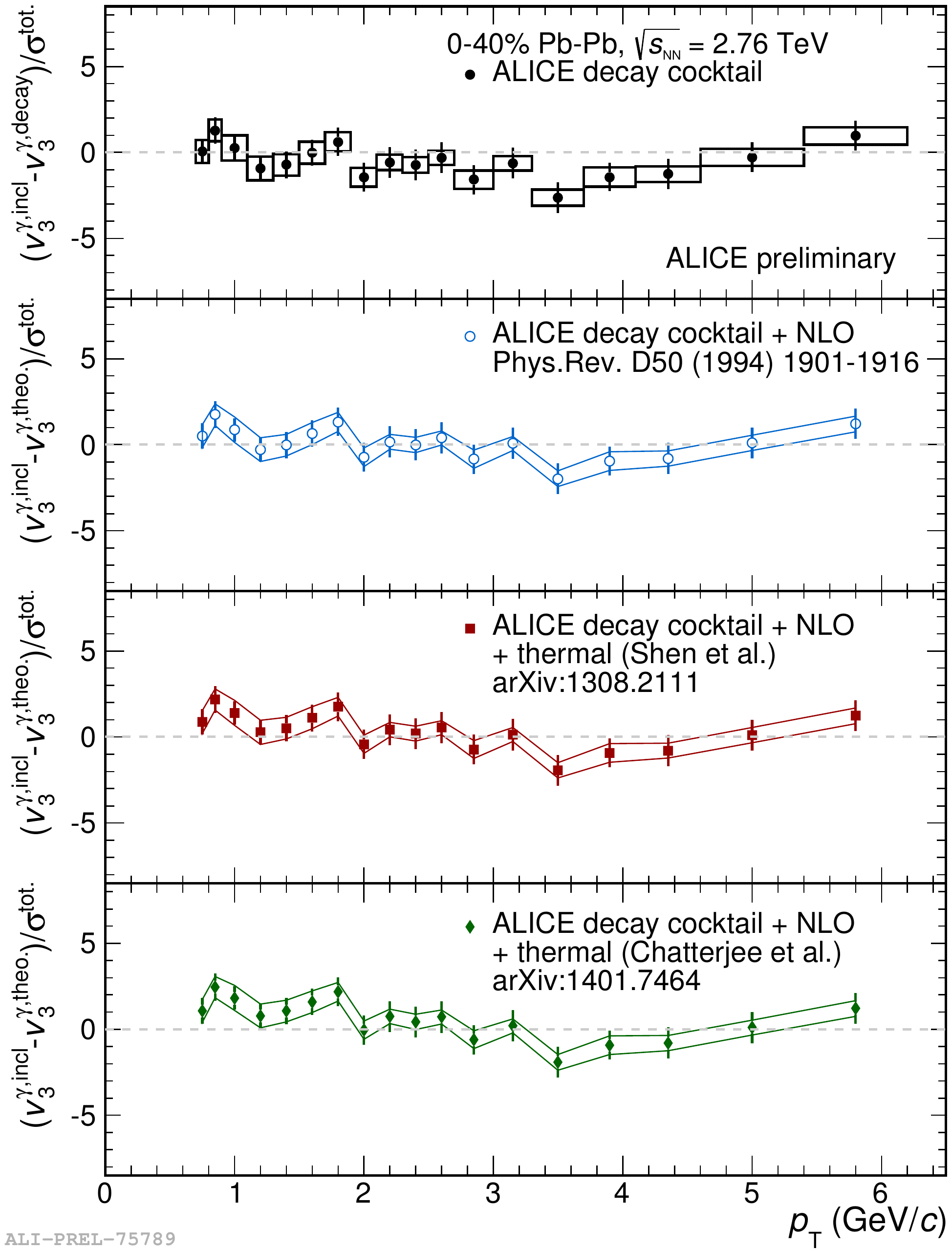}\figsubcap{b}}
  \caption{Same as fig.\ \ref{fig:v2} but for triangular flow.}%
  \label{fig:v3}
\end{center}
\end{figure}

The collective flow of direct photons is extremely interesting because it is
expected that direct photons are emitted from the hottest stage of the
collision and their flow reflects the development of collective expansion at
early stages. Experimentally, one can measure the flow of inclusive photons,
estimate the flow of decay photons and estimate the flow of direct photons as

\begin{equation}
  v_n^{\gamma, dir} = \frac{v_n^{\gamma, incl} R_\gamma - v_n^{\gamma, dec}}{1-R_\gamma} ,
  \label{vn}
\end{equation}

where $v_n^{\gamma, dir}$, $v_n^{\gamma, incl}$ and $v_n^{\gamma, dec}$ are the
flow of direct, inclusive and decay photons with respect to the $n^{\mathrm{
th}}$ harmonic. As one can see from equation \ref{vn}, with $R_\gamma$ close to
unity, the uncertainties rapidly increase. Therefore, we present the results of
direct photon flow as a comparison of collective flow of inclusive and decay
photons, see figs.\ \ref{fig:v2} and \ref{fig:v3} for elliptic and triangular
flow respectively. In fig.\ \ref{fig:v2}a (\ref{fig:v3}a) the elliptic
(triangular) flow contribution from inclusive photons is shown in red while the
decay photons are shown in black.  Predictions are also shown for
Next-to-Leading-Order (NLO) prompt photons plus decay photons (blue dash-dotted
line) \cite{Gordon:1994ut} as well as two models for inclusive photons (decay +
prompt + thermal) in the red \cite{Shen:2013cca} and green
\cite{Holopainen:2011pd} curves.  Quantitative comparisons are shown in the
right insets of these figures as well.  The differences are shown in units of the
sigma of the total uncertainties. We found that both elliptic and triangular
flow of inclusive photons are considerably smaller than the expected flow of
decay photons and agree with predictions of both models incorporating thermal
photon contributions.

\section{Conclusions}
We reviewed the ALICE results on production of neutral pions in pp collisions
at $\sqrt{s}=0.9$, 2.76 and 7 TeV and neutral pions and direct photons in Pb--Pb
collisions at $\sqrt{s_{NN}}=2.76$ TeV. QCD calculations reproduce the neutral pion
spectra in pp collisions at $\sqrt{s}=0.9$ TeV but overpredict the pion yields at
higher energies. In Pb--Pb collisions, neutral pion yields demonstrate a suppression
of a factor of $10$ in the most central collisions with respect to scaled pp
collisions at the same energy. The direct photon spectrum agrees with pQCD
predictions at $p_{\mathrm T}>4$~GeV/$c$ but shows excess at lower $p_{\mathrm
T}$. Collective flow of inclusive photons differs from estimates of collective
flow of decay photons and agrees with expectations including contributions from
thermal photons.

\section*{Acknowledgments}
This work was partially supported by the grant RFBR 12-02-91527.

\bibliographystyle{ws-procs975x65}
\bibliography{Peressounko}

\begin{thebibliography}{10}

\bibitem{Aamodt:2008zz}
K.~Aamodt {\em et~al.}, {The ALICE experiment at the CERN LHC}, {\em JINST}
  {\bf 3}, p. S08002  (2008).

\bibitem{Abelev:2014ypa}
B.~B. Abelev {\em et~al.}, {Neutral pion production at midrapidity in pp and
  Pb-Pb collisions at $\sqrt{s_{NN}}$ = 2.76 TeV}, {\em Eur.Phys.J.} {\bf C74},
  p. 3108  (2014).

\bibitem{Abelev:2012cn}
B.~Abelev {\em et~al.}, {Neutral pion and $\eta$ meson production in
  proton-proton collisions at $\sqrt{s}=0.9$ TeV and $\sqrt{s}=7$ TeV}, {\em
  Phys.Lett.} {\bf B717}, 162  (2012).

\bibitem{d'Enterria:2014zga}
D.~d'Enterria, K.~J. Eskola, I.~Helenius and H.~Paukkunen, {LHC data challenges
  the contemporary parton-to-hadron fragmentation functions}, {\em PoS} {\bf
  DIS2014}, p. 148  (2014).

\bibitem{Aad:2011fc}
G.~Aad {\em et~al.}, {Measurement of inclusive jet and dijet production in $pp$
  collisions at $\sqrt{s}=7$ TeV using the ATLAS detector}, {\em Phys.Rev.}
  {\bf D86}, p. 014022  (2012).

\bibitem{deFlorian:2014xna}
D.~de~Florian, R.~Sassot, M.~Epele, R.~J. Hernandez-Pinto and M.~Stratmann,
  {Parton-to-Pion Fragmentation Reloaded, arXiv:1410.6027}  (2014).

\bibitem{Adler:2006bw}
S.~Adler {\em et~al.}, {A Detailed Study of High-p(T) Neutral Pion Suppression
  and Azimuthal Anisotropy in Au+Au Collisions at $\sqrt{s_{NN}} = 200$ GeV},
  {\em Phys.Rev.} {\bf C76}, p. 034904  (2007).

\bibitem{Sharma:2009hn}
R.~Sharma, I.~Vitev and B.-W. Zhang, {Light-cone wave function approach to open
  heavy flavor dynamics in QCD matter}, {\em Phys.Rev.} {\bf C80}, p. 054902
  (2009).

\bibitem{Neufeld:2010dz}
R.~Neufeld, I.~Vitev and B.-W. Zhang, {A possible determination of the quark
  radiation length in cold nuclear matter}, {\em Phys.Lett.} {\bf B704}, 590
  (2011).

\bibitem{Horowitz:2011gd}
W.~Horowitz and M.~Gyulassy, {The Surprising Transparency of the sQGP at LHC},
  {\em Nucl.Phys.} {\bf A872}, 265  (2011).

\bibitem{Gordon:1994ut}
L.~Gordon and W.~Vogelsang, {Polarized and unpolarized isolated prompt photon
  production beyond the leading order}, {\em Phys.Rev.} {\bf D50}, 1901
  (1994).

\bibitem{Vog97a}
W.~Vogelsang and M.~R. Whalley, A compilation of data on single and double
  prompt photon production in hadron hadron interactions, {\em J. Phys.} {\bf
  G23}, A1  (1997).

\bibitem{Vog04a}
W.~Vogelsang. Private communications (August, 2004).

\bibitem{Shen:2013cca}
C.~Shen, U.~W. Heinz, J.-F. Paquet, I.~Kozlov and C.~Gale, {Anisotropic flow of
  thermal photons as a quark-gluon plasma viscometer, arXiv:1308.2111}  (2013).

\bibitem{Holopainen:2011pd}
H.~Holopainen, S.~Rasanen and K.~J. Eskola, {Elliptic flow of thermal photons
  in heavy-ion collisions at Relativistic Heavy Ion Collider and Large Hadron
  Collider}, {\em Phys.Rev.} {\bf C84}, p. 064903  (2011).

\end{thebibliography}

\end{document}